# Parametric analysis of Cherenkov light LDF from EAS in the range 30–3000 TeV for primary gamma rays and nuclei

© 2017 **A.Sh.M. Elshoukrofy**[a,b], **E.B. Postnikov**[a], **E.E. Korosteleva**[a], **L.G. Sveshnikova**[a], **H.A. Motaweh**[b]

[a] *Skobeltsyn Institute of Nuclear Physics, Moscow State University, Moscow, 119991 Russia*
e-mail: tfl10@mail.ru, evgeny.post@gmail.com

[b] *Faculty of Science, Damanhour University, El-Gomhouria St., 22516, Damanhour, El Beheria, Egypt*
e-mail: abeershehatamahmoud@yahoo.com

**Abstract** − A simple 'knee-like' approximation of the Lateral Distribution Function (LDF) of Cherenkov light emitted by EAS (extensive air showers) in the atmosphere is proposed for solving various tasks of data analysis in HiSCORE and other wide angle ground-based experiments designed to detect gamma rays and cosmic rays with the energy above tens of TeV. Simulation-based parametric analysis of individual LDF curves revealed that on the radial distance 20–500 m the 5-parameter 'knee-like' approximation fits individual LDFs as well as the mean LDF with a very good accuracy. In this paper we demonstrate the efficiency and flexibility of the 'knee-like' LDF approximation for various primary particles and shower parameters and the advantages of its application to suppressing proton background and selecting primary gamma rays.

## INTRODUCTION

The primary gamma-ray flux follows a power law and rapidly decreases with energy. Therefore, in the energy region beyond 10 TeV large effective area arrays (multi km$^2$) are required. One of these arrays is the TAIGA complex hybrid detector system [1, 2], which consists of the HiSCORE (High Sensitivity Cosmic ray Origin Explorer) [3] wide angle Cherenkov timing array and several imaging air Cherenkov telescopes. Previous wide angle timing array experiments for gamma-ray astronomy were realized on a smaller scale, e.g. the AIROBICC instrument [4] had an area of less than 0.1 km$^2$ and was optimized for gamma ray energy <10 TeV. The methods of data analysis in HiSCORE are based on those developed for the Tunka-133 experiment [2]. They implement a Cherenkov timing-array technique designed for cosmic ray nuclei study in the PeV–EeV range.

Gamma-ray astronomy measurements require high energy resolution for spectroscopic studies, high directional resolution for pinpointing the location of the observed sources, and strong hadron background suppression [5], because a sensitivity of the telescopes depends linearly on the latter two factors. The wide angle technique assumes that the air shower front timing is measured by counting the number of Cherenkov photons emitted by secondary air shower particles and subsequently registered in every station of the array. As the time front has a cone shape instead of a plane one, the accuracy of arrival direction reconstruction depends linearly on that of a shower core position. The goal of this paper is to study knee-like fitting functions of Cherenkov light LDF for gamma rays and charged cosmic ray nuclei at the energy interval of 30–3000 TeV, which has never been covered before. In this paper we studied the problem of gamma ray event selection and cosmic ray background rejection. The following tasks: shower core position and energy reconstruction, directional reconstruction, and the mass group identification of cosmic rays, will be covered in the next paper.

## KNEE-LIKE APPROXIMATION

We performed our investigation on CORSIKA-simulated [6] data of the number of Cherenkov photons, $Q(R)$, emitted by showers in the atmosphere and detected with a space step of $R$=5 m in the shower frame perpendicular to the shower direction. Simulation was performed for the zenith angle range 0–50° for primary gamma rays, protons, and nuclei (He, C, Fe) with the energy $E$=30–3000 TeV. For parameterization of the simulated $Q(R)$ for every event, we propose a simple function designated as a 'knee-like approximation', which was used earlier by J. Horandel [7] as a function of energy for description of the 'knee' in the cosmic ray spectrum. In our approach we describe the radial density of Cherenkov photons $Q(R)$ for individual events. It depends on five parameters $C$, $\gamma_1$, $\gamma_2$, $R_0$, and $\alpha$:

$$Q_{appr} = CR^{\gamma_1}(1+(\frac{R}{R_0})^\alpha)^{\frac{\gamma_2}{\alpha}} \qquad (1)$$

In Fig. 1 we plot simulated individual $Q(R)$ distributions for protons, helium, and gamma rays with the energy 30 and 100 TeV. In spite of the significant diversity, for all the events the LDF has a specific knee-like structure. A flat direct Cherenkov light disk ($Q(R) \sim R^{\gamma_1}$, with the slope $\gamma_1 \sim -0.7 \div 0.2$) is extended up to the distance $R_0$ (a 'knee' position). In the region beyond $R_0$, $Q(R) \sim R^{\gamma_1+\gamma_2}$ and a photon density decreases very steeply with the slope $\gamma_1+\gamma_2 \sim -2$, this part represents a Cherenkov radiation caused by multiple scattered electrons. The $R_0$ value depends strongly on the distance to the shower maximum and it ranges from 75 to 200 m. The parameter $\alpha$ is responsible for the sharpness of the knee [7] and provides a smooth transition between the power laws at the knee point. This parameter is the main distinction between our approximation and the ones proposed in the previous works. It allows us to describe both cases: a very smooth transition from the central part to the periphery across the knee (see Fig. 1 *b*), in this case α<3, and a very sharp transition (see Fig. 1 *e*), in this case usually α>50. The value of α ranges from 0 to ∞, and its logarithm can classify the LDFs as belonging to 2 separate clusters: smooth ones and sharp ones, as will be seen later in Fig 2 *c*.

Previous efforts to approximate Cherenkov light LDF had various limitations on the fitting accuracy and/or the conditions of application. In [4, 5] piecewise fitting functions are defined on two intervals and the knee point is a derivative jump discontinuity. In the Tunka-133 experiment a piecewise function [8, 9], composed of 4 pieces, was designed to fit both smooth and sharp knee-like LDFs. The total number of parameters of all constituent functions was reduced to two: *a* and *bxy*, the first one is a normalization factor, the second one characterizes the steepness of the whole LDF. The Tunka-133 approximation well describes all the LDFs for primary cosmic rays in the energy interval $10^{15}$–$10^{18}$ eV, for which it was elaborated. Nonetheless, our analysis reveals that the LDF of gamma rays, especially incident at large angles, very often has a positive value of $\gamma_1$, which the Tunka-133 fitting function cannot reproduce. The two-parameter fit of Tunka-133 has a distinctive advantage when we work with a small number of triggered detectors, however it reduces the diversity of the LDFs to a fixed set of curves. More complicated LDF approximations for Cherenkov light from EAS were proposed in [10, 11], they are also optimized for cosmic rays at PeV energies, but not for *sub* TeV gamma rays.

The knee-like approximation (1) was applied to all individual events from the HiSCORE simulation bank for various types of primary particles at different energies (in Fig. 1 the discrepancy between the simulated $Q(R)$ and the approximations is indistinguishable). In the Table we present the mean values of parameters for 3 types of primary particles and 6 values of energy from 30 to 3000 TeV at two intervals of azimuth angles. Parameter dependence on energy, angle, and type of primary particle is clearly visible. Analysis shows that the new approximation function fits the individual LDFs as well as average LDFs with a very good accuracy: the mean squared error of the fit ~ 0.0005–0.002 in logarithmic scale for all energies and for each type of particle. It allows us to obtain energy dependencies, to study correlation between parameters, and to develop parametric methods of primary particle identification.

## APPLICATIONS

New approximation can be used for a variety of tasks in the HiSCORE experiment: shower core position and energy reconstruction, background rejection and gamma ray selection, separation of primary light nuclei and heavy ones. Solving most of these problems will be described in the next paper, because it requires the real experimental conditions to be considered. In Fig. 2 the distributions of 3 parameters ($\gamma_1$, $\gamma_2$, lg$\alpha$) are presented for true LDF functions, which were simulated with the same spacing of 5 m as for the previous Fig. 1, for 100 TeV protons and gamma rays. For all of these parameters both gamma ray and proton distributions are separated from each other, and this fact can be used to discriminate between sorts of particles. For example, the discrimination between gamma rays and protons in a 3-dimensional space of these parameters using quadric separating surfaces in quadratic discriminant analysis [12] leads to strong proton suppression by a factor as large as ~100. In the next paper this procedure will be adapted to the real experimental conditions.

## CONCLUSIONS

We proposed a simple 'knee-like' approximation of Cherenkov light radial distribution emitted by EAS and tested the quality of these approximations. The new approximation gives the possibility to fit the whole diversity of individual LDFs for different nuclei and gamma rays on the shower core distance 20–500 m at the energy interval 30–3000 ТэВ with a very good accuracy. Parameters of approximation $\gamma_1$, $\gamma_2$, $R_0$, and $\alpha$ depend on the energy and the type of primary particle and allow us to separate proton and gamma ray induced showers: 3-parameter multivariable approach leads to a strong suppression of background. Application of this fit to the real experimental conditions is to follow in the next paper.

The study was supported by the Russian Science Foundation, project no. 15-12-20022.

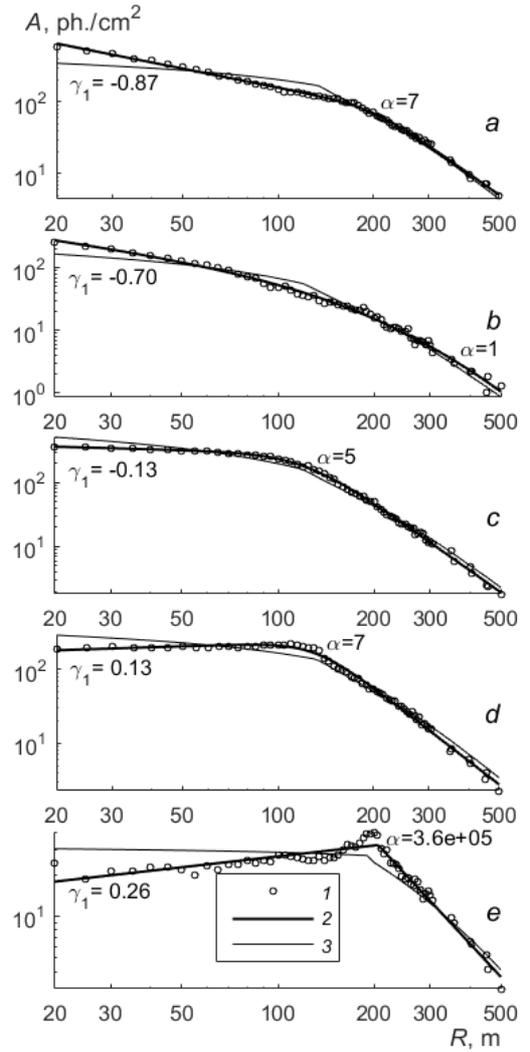

Fig. 1.
Five individual simulated LDFs of the photon pulse amplitude:
*a* – helium 100 TeV at zenith angle Θ=43.7°,
*b* – proton 30 TeV at Θ=48.5°,
*c–e* – gamma 30 TeV at Θ=26.8°, 15.2°, 49.6°;
*1* – true LDF,
*2* – knee-like fit,
*3* – Tunka-133 fitting function

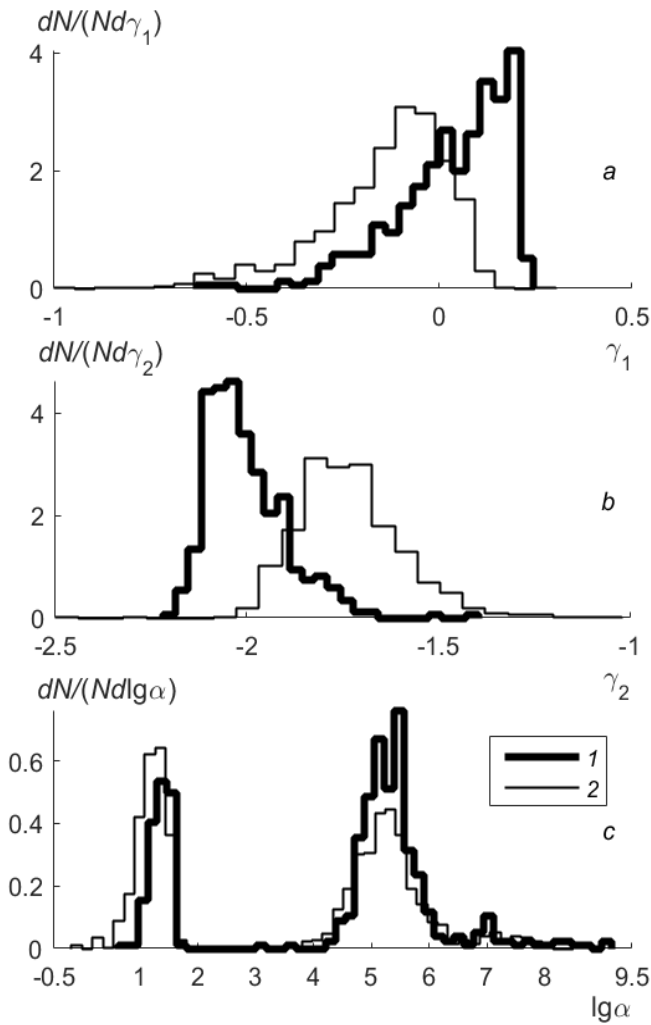

Fig. 2.
Comparison of the fitting parameter distributions for protons and gamma rays with the energy 100 TeV at zenith angles 25°<Θ<50°:
*a* – $\gamma_1$,
*b* – $\gamma_2$,
*c* – lg$\alpha$;
*1* – gamma rays,
*2* – protons.

Table. Mean values of $\gamma_1$, $\gamma_2$, $R_0$, and $\alpha$ in the approximation function (1) obtained for 3 types of primary particles and 6 values of energy for various ranges of azimuth angle $\theta$.

| Parameters | $\gamma_1$ | $\gamma_2$ | $R_0$ | $\alpha$ | $\gamma_1$ | $\gamma_2$ | $R_0$ | $\alpha$ |
| --- | --- | --- | --- | --- | --- | --- | --- | --- |
| Type | $\theta<25°$ | $\theta<25°$ | $\theta<25°$ | $\theta<25°$ | $\theta>30°$ | $\theta>30°$ | $\theta>30°$ | $\theta>30°$ |
| Pr 30 TeV | −0.26 | −1.69 | 114 | 27.9 | −0.118 | −1.665 | 150.6 | 55 |
| Pr 100 TeV | −0.37 | −1.63 | 110 | 24 | −0.137 | −1.73 | 148 | 49.7 |
| Pr 300 TeV | −0.43 | −1.64 | 106 | 21 | −0.179 | −1.72 | 144 | 42.8 |
| Pr 1000 TeV | −0.48 | −1.66 | 102.8 | 20 | −0.24 | −1.709 | 135.7 | 35 |
| Pr 3000 TeV | −0.61 | −1.54 | 96 | 19 | −0.28 | −1.69 | 132 | 29 |
| | | | | | | | | |
| Gam 30 TeV | −0.125 | −1.90 | 110.2 | 50 | 0.13 | −2 | 154 | 71 |
| Gam 100 TeV | −0.25 | −1.83 | 104 | 25 | 0.07 | −2 | 148.9 | 62.5 |
| Gam 300 TeV | −0.34 | −1.78 | 99.6 | 22 | 0.01 | −1,98 | 144 | 55 |
| Gam 1000 TeV | −0.41 | −1.67 | 95 | 20 | −0.13 | −1.89 | 130 | 35 |
| | | | | | | | | |
| Fe 100 TeV | −0.20 | −1.64 | 127.1 | 12 | −0.12 | −1.56 | 167 | 14 |
| Fe 300 TeV | −0.28 | −1.57 | 120.4 | 13.4 | −0.16 | −1.52 | 160 | 21 |
| Fe 1000 TeV | −0.37 | −1.55 | 115.6 | 21 | −0.21 | −1.58 | 148 | 35.6. |